\documentclass[epsfig,12pt]{article}

\usepackage{mathrsfs}
\usepackage{latexsym}
\usepackage{amsthm}

\usepackage{graphicx}
\usepackage{amsmath}
\usepackage{makeidx}
\usepackage{color,graphics,palatino}
\usepackage{amsfonts}
\usepackage{amssymb}
\usepackage{epsfig}
\usepackage[all]{xy}
\makeatletter \@addtoreset{equation}{section}

\def\be{\begin{equation}}
\def\ee{\end{equation}}
\def\bea{\begin{eqnarray}}
\def\eea{\end{eqnarray}}

\newcommand{\nc}{\newcommand}
\nc{\al}{\alpha} \nc{\bib}{\bibitem} \nc{\la}{\lambda}
\nc{\C}{\mbox{\hspace{1.24mm}\rule{0.2mm}{2.5mm}\hspace{-2.7mm}
C}} \nc{\R}{\mbox{\hspace{.04mm}\rule{0.2mm}{2.8mm}\hspace{-1.5mm}
R}}
\begin{document}\title{%
\rightline{\mbox {\normalsize {LPHE-MS-11-02/
CPM-11-02}}\bigskip}\textbf{  The Yang Monopole in IIA
Superstring:\\ Multi-charge Disease and Enhan\c con Cure   }}
\author{Adil  Belhaj$^{1,2,5}$\thanks{belhaj@unizar.es}, Pablo  Diaz$^{3}$\thanks{pablo.diazbenito@wits.ac.za},
Antonio  Segui$^{4}$\thanks{segui@unizar.es}\\
{\small $^{1}$Lab Phys Hautes Energies, Modelisation et Simulation, Facult\'{e} des Sciences, Rabat, Morocco} \\
{\small $^{2}$Centre of Physics and Mathematics, CPM-CNESTEN, Rabat, Morocco } \\
{\small $^{3}$National Institute for Theoretical Physics, University
of Witwatersrand, South Africa}\\
{\small $^{4}$Departamento de Fisica Teorica, Universidad de Zaragoza, 50009-Zaragoza, Spain}\\
 {\small $^{5}$Groupement National de
Physique des Hautes Energies, Si\`{e}ge focal: FSR, Rabat, Morocco }
} \maketitle

\begin{abstract}
\bigskip
A brane picture in Type IIA superstring  for the Yang Monopole is
reconsidered. It makes use of
 D2 and D4-branes wrapped on cycles  in the K3  surface. When the model
was first presented some problems concerning the charges of the monopoles arised. In this paper, they are
shown to be
 cured by the model itself. Surprisingly, the incompatibility between the multi-charge configuration and
the spherical symmetry of the Yang Monopole is seen in the brane
description as the
 emergence of the enhan\c con shell and the fuzzy geometry.
This consistency is deep and surprising, and is the point that triggered this work.
 It nontrivially relates a purely geometrical problem in
ordinary spacetime with the emergence of noncommutative geometries.
Besides, this paper includes an extended model for $SO(4)$-monopoles, a T-dual model
 in Type IIB superstring  and an analysis  on the possible duality  between our model
and another setup in M-Theory/Heterotics for the Yang monopole found
before. \\\textbf{Keywords}: Yang Monopole, Supertring theory,
Dualities, M-theory,   K3 surface.
\end{abstract}
\newpage
\tableofcontents

\newpage

\newpage
\section{Introduction}


Based on the idea that string theories at low energies must be well
approached by supergravity, we have been searching for a brane model
in type IIA string theory of the Yang monopole~\cite{BDS2}. In the
brane picture shown in~\cite{BDS2} most of the features of the Yang
construction where successfully reproduced, namely the charges, the
nonabelian field and the point-like behavior of Yang configurations.
Some corners of the brane model remained unanswered, however,
questions whose investigation have revealed some unexpected deep
connections between gauge theory and brane physics and have finally
lead to this paper.

Before going deeper into this
let us recall some general details of the Yang monopole.

As a pure gauge solution of  the Yang monopole~\cite{Y} was first
constructed as a generalization of the Dirac monopole~\cite{D}.  The
Yang configuration  is characterized by the flux of the four-form
field $Tr F \wedge F$, charged under
 the  $SU(2)$   gauge group, across the four dimensional sphere
 that covers the origin in a $5+1$
 dimensional
 space-time. It corresponds to the conformal mapping into $S^4$
 of the BPST  Euclidean
 instanton solution~\cite{BPST}.  Again,  the origin  is singular
 but now the energy of this solitonic configuration
is well behaved in the UV regime, although IR  divergences
linearly appear. The total energy
 inside a four sphere
  is proportional to  its  radius. In this  realization,
  the flux is quantized but now, the magnetic charge
   of the Yang monopole
 can take only  two values $\{+1,-1\}$~\cite{Y}. This charge,
 which may    correspond to
   the self-dual and anti-self-dual BPST  instanton
 configurations respectively, is given by the integral
 over $S^4$ of the second Chern
  class $Tr(F \wedge F) /8 \pi^2$. The Yang monopole can
  be easily generalized to higher  even
dimensional space-time  where  the configuration is
characterized by
$2n$-form $F^n$. Explicit solutions can be systematically obtained in
\cite{HP,Tc,GT,DS}.

 We would like to stress at this point the well known fact that the Yang monopole,
  {\it due to its spherical symmetry},
can carry just two charges~\cite{Y}. This is a purely gauge requirement in the
sense that, unlike the {\it fuzzy} backgrounds which emerge in the brane picture, the charge
analysis from the gauge theory point of view concerns topologically nontrivial configurations
 of gauge fields in an {\it ordinary}
4-dimensional sphere.  Indeed, in the process of generalizing the Yang construction for larger
 symmetry groups and dimensions,
there is a systematic way of finding the number of configurations (charges)
 that an arbitrary gauge field with symmetry group $G$ on a $K/H$ symmetric (ordinary) base space
can show once $K$-symmetry on the field configuration is
imposed~\cite{DL}. The configurations over a symmetric space are
labeled by the so-called Wang maps. Wang maps are homomorphisms
between $H$ (the isotropy group) and the symmetry group of the gauge
field $G$ up to isomorphism. Note, again,
 that the whole analysis in the general case in~\cite{DL} involves an {\it ordinary} spacetime as a base space.
  Note, also, that when we
talk about spherical symmetry we refer to the case in which the
symmetric base space is a sphere. For the Yang case, the base space
is $S^4\subset R^{5,1}$, the $K$-symmetry is
 $SO(5)$\footnote{The symmetry
is actually $spin(5)$, the double cover of $SO(5)$ as proved
in~\cite{DL}, but this subtlety is not going to make much difference
in our discussion.}, and the gauge group $G$ is $SU(2)$.

In the brane model presented in~\cite{BDS2} the two charge-nature of
the Yang monopole seemed to hold nicely as associated with the two
ways a single D4-brane can wrap a 2-cycle of the K3 surface.
Specifically, the construction given in ~\cite{BDS2} needs D2 and
D4-branes on the K3 surface. In this brane picture, the $SU(2)$
 gauge  group of the Yang construction is {\it engineered} by means of
 a D2-brane wrapping
 shrinking 2-cycles inside  the  K3 surface. The  Yang monopole  comes  up when
 a D4-brane wraps  the whole K3 and dimensional reduction to
 6-spacetime dimensions is performed. In this way, the Yang properties
 above mentioned are  encoded in
the K3 surface features. However, a careful look reveals a mechanism
that allows us to obtain a multi-charge spherically symmetric
configuration. Basically, the mechanism consists of adding
subsequent D4-branes to the setup at no cost of energy since the
branes are taken to be BPS states. The addition of $N$ D4-branes
would increase the charge in $N$ units without loosing spherical
symmetry since the branes are point-like in 5+1 dimensions. It would
lead, as said, to a multi-charge spherically symmetric
configuration.

At this stage, the brane setup would be ruined as a model for the
Yang monopole given that, as said before, no spherical SU(2)
configuration over $S^4$ can be multi-charged. However, a deeper
investigation on the brane physics reveals the appearance of an
enhan\c con shell~\cite{JPP,JMPR,DPPR,FLM,AM,J,J2, CMT} and a fuzzy
geometry at the core of the monopole for the multi-charge case.
Fuzzy geometry explicitly breaks spherical symmetry. So the
two-charge property of the spherically symmetric gauge theory
configuration gets restored. The surprise comes because spherical
 symmetry is broken in the brane picture
by a {\it non-ordinary} spacetime background: the (noncommutative) fuzzy 4-sphere that the enhan\c con
mechanism brings aside.
In this highly nontrivial way, which involves fuzzy geometries, a gauge requirement for the two allowed
charges of the Yang configuration is recovered in the brane picture.


Apart from the discussion of the multicharge configurations and the
broken spherical symmetry, this paper presents a new brane model
for $SO(4)$-monopoles in six dimensions. The construction is
inspired in the model presented in \cite{BDS2}. It is, say,
its natural extension. The number of charges labeling the
topologically different brane configurations in this case
is 4. Needles to say that the same argument used for the
Yang monopole holds for the $SO(4)$ construction and also
prevents it from having an infinite tower of charges
without breaking spherical symmetry.

The organization of this paper is as follows.  In section
\ref{sec:TSCYM} we will review the model proposed in~\cite{BDS2},
discuss the relation with the M-theory/Heterotic string setup and
find a useful T-dual version of the Type IIA model in Type IIB
superstring. In section \ref{sec:so4} we will extend the logic of
the model for the case of $SO(4)$-monopoles. Section \ref{sec:MCDEC}
deals with the core of the discussion. It is devoted to the
discussion of the main objections concerning the multiple charge
configurations the model presented at first sight. There will be
offered, to these objections, a solution coming from the model
itself by a purely stringy effect: the enhan\c con mechanism and the
fuzzy geometry which comes along with it. Both, intriguing
phenomena, which are better visualized in the T-dual model.
 Finally, a brief conclusion sums up the
main points of the paper and brings some open questions.

\section{Type IIA  superstring   construction  of the Yang monopole}\label{sec:TSCYM}
The idea of finding a brane picture for Yang monopoles is not new
though. Before~\cite{BDS2}, it was suggested the possibility of
considering the Yang Monopole in  M-theory~\cite{P}.  In this
regard, a  heterotic M-theory  realization was soon
proposed~\cite{BGT}\footnote{ Independently, a matrix model of the
Yang  monopole was given in ~\cite{CIK}.}. In  particular,
\cite{BGT} shows  that M5-brane may have boundaries on M9-branes,
where the  boundary is  a D4-brane with an infinite tension so its
centre
 of mass is not free  to move.  The  boundaries  may be identified  with  Yang monopoles.
 Indeed, in this  heterotic M-theory picture there are  two Yang monopoles which correspond  to the ends of
the oriented M5-brane which stretches between two M9-branes. Each
monopole (each end) is charged under an $SU(2)$ subgroup  of  $E_8$
with the  topological charges  $\{ +1,-1\}$ respectively. Using
string/string duality
  and the  result of~\cite{BGT},  we
give  a string   realization   of the Yang monopole
  for  a  six dimensional  Type IIA superstring obtained
  from the compactification on  a  local description of the  $K3$  surface
  in the presence of  wrapped  D-branes.  Then  we  study its relation with the
M-theory model.
\subsection{ Superstringy   construction  of the Yang monopole}\label{sec:SCOTYM}
 Consider a local description of the K3 surface
 where the manifold
develops a $su(2)$
  singularity (known as  $A_1$ singularity).  This singularity
corresponds to a  vanishing   two-sphere.   Near  such a singular point,
  the K3 surface    can be identified
 with   the asymptotically  locally  Euclidean   (ALE)   complex space  which
  is algebraically given by
\begin{equation}
 f(x,y,z)=xy-z^2=0, \label{A1}
\end{equation}
  which  is  singular  at $x=y=z=0$.  In two-dimensional $N = 2$  linear
 sigma model with only one\footnote{For the ALE geometry $A_n$ the gauge group is
 $U(1)^n$.} $U(1)$ gauge symmetry,   the resolution of this singularity  is related to the
 D-term described  by the following bosonic potential $V(\phi_1, \phi_2,
 \phi_3)$
\begin{equation}
V(\phi_1, \phi_2, \phi_3) = ( |\phi_1|^2- 2|\phi_2|^2+|\phi_3|^2)-R)^2,
\end{equation}
  where  $R$  is the $U(1)$  Fayet-Iliopoulos (FI) parameter\footnote{In this way,
one sees that the $U(1)$  Cartan subgroup of the $SU(2)$  symmetry of the singularity of $K3$
 carries the gauge symmetry
of the $N = 2$  supersymmetric linear sigma model.}~\cite{W}. Geometrically,
this corresponds
 to replace the
singular point $x = y = z = 0 $ by  the 2-sphere $S^2$ defined by $
V=\phi_2=0$,
 which  is  the  only  non-trivial 2-cycle   on which we can
 wrap D2-branes.   In order to  geometrically engineer the
 SU(2) gauge symmetry
 only the compact  piece containing  the  $S^2$  is
 necessary~\cite{KKV}. Now, the system consists
  of    Type IIA  D2-branes wrapping around     $S^2$.  This   gives
 a pair of massive  vectors $W^{\pm}$,  one for each of
the two  possible   ways   of the wrapping.
  The  masses of these particles are proportional to  the volume of  the  2-sphere.
   They are charged under  the  $U(1)$  gauge  field
   obtained by decomposing the type IIA
 three-form in terms of the  harmonic 2-form  on the  2-sphere  and the
  1-form gauge field   in  six dimensional space-time. In the
   limit where  the 2-sphere shrinks,
the $W^{\pm}$ particles become  massless and, together
with the one form gauge field,
  generate  the $SU(2)$  adjoint representation.  This
  will be  identified with
the  gauge symmetry of our Yang monopole. We have obtained the
electrically charged sector, associated to D2-branes wrapping two
 vanishing  cycles in
the  K3 surface. Lifting  consistently to 11 dimensions,  the M2-brane  is
encountered.  The
 magnetic Yang
monopoles can be identified  with  D4-branes,  totally
wrapped   on  the K3 surface. As consequence, they  generate the magnetic
 objects in   the six dimensional space-time. This
 is expected from the fact that the
  D4-brane
 is the only magnetic object in  Type IIA superstring
 theory which  can be obtained from
   the M5-brane  and gives a
point-like particle  after wrapping the K3 surface.
Owing to the spherical symmetry
 of the six dimensional configuration and on the fact
 that, as seen above, the gauge group
  origin  is linked to  the singular  limit of the
  geometry, we strongly believe that  all
    Yang monopole  properties should
be derived from the K3 surface  data.

Schematically, the ten dimensional spacetime where Type IIA lives is occupied as
follows. If we consider that the K3 surface extents along dimensions 6789, and the
vanishing 2-cycle of K3 at 67 positions, then the D2-brane is at 067 and the D4-brane at
06789.

 We will show that the
charges $\{+1,-1\}$    can have different    compatible  K3 surface interpretations.
First, the different ways in which D4-branes  are wrapped
on  the K3 surface. They  are classified
 by its  fourth homotopy group. As seen before, in order
 to construct the $SU(2)$ gauge
  group, it is necessary to work with a local model of K3  with a
  singularity  $A_1$. The deformed
   geometry is   given  by the product of the complex $ {\bf C}$
   plane and a two sphere $S^2$.
    Since $\Pi_q(X \times Y)=\Pi_q(X )\times \Pi_q(Y)$,  we
    have the following remarkable relation
 \begin{equation}\label{eq:homo}
 \pi_4(A_1)\sim\pi_4(S^2)=Z_2.
\end{equation}
 The two charges of the Yang monopole are related to the  two ways the geometry allows
 a D$4$-brane to wrap on  it\footnote{This last statement
 has left some room for controversy. After reading our
paper~\cite{BDS2}, David Tong suggested that there should
be five charges corresponding
 to the two ways the D4-brane and the D4-antibrane wrap
 the $A_1$ manifold plus the
 trivial one (zero charge). We believe that
 David's idea is right for the case of
  extended-Yang monopoles,
   the ones with gauge group $SO(4)$, we will go on
   this point in subsection \ref{sec:so4}.}.

It is known that the energy of the Yang monopole diverges linearly in spacetime.
 This fact is not manifest in this geometric construction. However, the divergence
in the energy can be intuitively seen in the T-dual model (see sec. \ref{DMTIIB}), where
the D4-brane turns into an infinite effective D1-brane which pulls at a pair of
coincidents NS5-branes.

  \subsection{Relation with the heterotic M-theory Yang monopole
configuration}\label{sec:RWTHMTYMC} The authors of~\cite{BGT} have
suggested a
  Yang monopole  representation
with two $SU(2)$ gauge  factors obtained by breaking the  $E_8
\times  E_8$  heterotic
 gauge symmetry in ten
 dimensions. This breaking, $E_8 \to E_7\times SU(2)$,
 could be related with the fact
 that the extremes of the M5-branes
 are located on the M9-branes and they  are just
 the core of the Yang monopole.  On the
  Type IIA side however   there is only one $SU(2)$
factor  which comes from  a D2-brane  wrapped around the collapsing
$S^2$  inside the K3 surface. Lifting to  M-theory the nature of
 this difference is appreciated.
Reduction  from 11 to 6 dimensions with sixteen supercharges  can be
performed  in two dual
  ways depending on the action
 of the $Z_2$ symmetry on the five dimensional internal space $S^1\times T^4$.
 In  the heterotic realization
 of M-theory, the $ Z_2$ symmetry acts on the  $S^1$
 factor giving rise the segment between
 the two M9-branes,
while in   the  Type IIA, M-theory,   the symmetry acts on  the
$T^4$ factor producing the $K3$ geometry. However,  since  these two
M-theory compactifications
 are dual in six dimensions,
  the  two above string  Yang  monopole realizations
 should be connected.  In what follows,  we will speculate on  this
 link. Indeed, the K3 surface   has two  possible
 constructions as the target space
of a sigma model.
 They  depend on the R-symmetry
 of the supercharges. Previously, we have mainly
 concerned with   $N=2$ sigma model,
  where the R-symmetry
 is supported by a $U(1)$ group, and the $K3$ target
 space  gets  manifested as a Kh\"{a}ler manifold.
  Now,  let  us   use the
 other realization of $K3$ where the manifold
 is hyperkhaler and the corresponding sigma model involves
 eight supercharges and  has a $SU(2)$ R-symmetry.
 Then, the  $\{+1,-1\}$  charges of the Yang monopole
 can  be explained by
 physical arguments  when the K3 surface is constructed
 in terms of   $N = 4$ sigma model~\cite{AW,B}.   This
  is  related to the   heterotic M-theory configuration
  where  the Yang monopole has two
    copies in the boundaries
  of the  M5-brane suspended between  two  M9-branes.
Theses copies, with charges $+1$ and $-1$,   might  be understood as
two hypermultiplets
 appearing  in the hyper-Kh\"{a}ler quotient construction of the $A_1$  space.

The six dimensional $SU(2)$   Yang Mills   theory can  also be
obtained from the  K3 surface that is realized in terms of $N = 4$
supersymmetric sigma model. This sigma model has only one
 $ U(1)$  gauge group,   two  hypermultiplets   with
 charges  $(q_1,q_2)$, and  one isotriplet FI
 coupling ${\vec \xi} = (\xi_1, \xi_2, \xi_3)$
 ~\cite{AW,B}. The sigma model gauge symmetry
 is related  to  the Cartan subgroup of the six
   dimensional gauge group.  In this construction,
the K3 surface is expressed by the vanishing condition of the
following D-terms
\begin{equation}
\sum_{i=1}^2q_i( \phi_i^\alpha
{\bar\phi}_{i\beta}+\phi_{i\beta}{\bar\phi}_i^\alpha)-{\vec
\xi}{\vec \sigma}^{\alpha}_{\beta}=0. \label{hyper}
\end{equation}
The double index $(i, \alpha)$  of the scalars refers to  the
component field doublets ($\alpha$) of the  two  hypermultiplets
($i$), and $\sigma$ are the traceless $2 \times 2$ Pauli matrices.
The condition under which the gauge theory flows in the infrared to
2d $N = 4$  superconformal field theory, which   is also the
condition to have a
 hyper-Kh\"{a}ler
Calabi-Yau background, is
\begin{equation}
q_1+q_2=0.
\end{equation}
This equation has different solutions that can be seen as
redefinitions of the coupling constant $\vec \xi$. Due to its
conformal invariance, the theory does not get affected by
redefinitions of $\vec \xi$, so the charge can be fixed to $-1$ and
$+1$.

Let us discuss the construction of the $K3$  surface in this case.
The starting point consists
 of two hypermultiplets with four scalars each.
 They can be expressed as ${\bf R^4}\times {\bf R^4}$.
The gauge invariance of each hypermultiplet (with $+1$ and $-1$
charge) under the
 $U(1)$ symmetry, together with the invariance under the $SU(2)$ R-symmetry that
  rotates the supercharges, enables us to express
  the  $K3$ surface locally as the following
   homogenous space
\begin{equation}
\frac{{\bf R}^4 \times {\bf R}^4}  {U(1)\times SU(2) }.
\end{equation}
There is a $Z_2$ symmetry that interchanges the two hypermultiplets
(the two ${\bf R}^4$ factors). We interpret the two ${\bf R}^4$
factors with their corresponding charges  as the two Yang monopole
copies which are the boundaries
 of the M5-branes  on the two M9-branes.
At this stage, we can ask the following question, which is the role
of the $SU(2)$ R-symmetry in this construction? The answer of this
question lies on the association of the R-symmetry with the
 instantonic nature of the M5-brane in the context of the heterotic M-theory picture.

\subsection{Six dimensional effective field theory}\label{sec:SDEFT}
The six dimensional field theory that remains after compactification of the IIA
supergravity and ignoring all massive Kaluza Klein modes can be consistently truncated
to:

\begin{equation}\label{effectiveaction}
S=\int{\sqrt{-g}\Big(R-(\partial\sigma)^2-e^{-\sigma}\text{tr}|F|^2\Big)},
\end{equation}
where $\sigma$ is the 6d dilaton and the trace is taken over the
color indices of the $SU(2)$ gauge field. Monopole solutions from
(\ref{effectiveaction}) are obtained with a spherically symmetric
ansatz for the metric:
\begin{equation}\label{sphericalmetric}
ds^2=-e^{\lambda(r)}\Delta(r)dt^2+dr^2/\Delta(r)+r^2d\Omega_4^2,
\end{equation}
in terms of functions $\lambda$ and $\Delta$, and $d\Omega_4^2$ which is the invariant
metric over the 4-sphere, together with the Yang monopole field strength and radial
ansatz $\sigma(r)$ for the dilaton.

The dilaton cannot be consistently set to a constant and then
eliminated from the action. So the solutions which minimize the
action (\ref{effectiveaction}) are in principle different from those
of a pure Yang-Mills theory to which the Yang monopole belongs.
However, because the Yang-Mills field strength 2-form for Yang
monopoles solution has components only on the 4-sphere, it continues
to minimize the action (\ref{effectiveaction}) and solve the
Yang-Mills equations as modified by the dilaton. This is the reason
why we still keep the name.

\subsection{Dual model in Type IIB superstring}\label{DMTIIB}
As we will see in the following, the appearance of the enhan\c con mechanism and the fuzzy geometry in the realization of
the Yang monopole we describe in this paper is better visualized in a dual model on Type
 IIB side. In section \ref{sec:RWTHMTYMC} we related our D2-D4 system on K3 setup with the
M-Theory (and heterotic string theory) model proposed in~\cite{GT}
by S-duality. As already noticed in~\cite{JPP}, a configuration with
$N$ D$(p + 1)$-branes stretched between two NS5-branes is T-dual to
the same number of D$(p + 2)$-branes wrapped on the two-cycle of an
$A_1$ ALE space, giving rise to $N$ effective D$p$-branes. This
means that the D2-D4 brane on a vanishing 2-cycle of K3 setup is
dual to a D1-brane and a D3-brane stretched between two approaching
NS5-branes, the vanishing limit of the IIA picture corresponding to
the coalescence of the NS5-branes. The spatial position of the
branes in the IIB picture can be described as follows. Let us state
that both NS5-branes occupy dimensions 012345. The D1-brane extends
along a transversal dimension (say, 6) and has an end on each
NS5-brane. The D3-brane, which is actually supporting the magnetic
charge of the configuration, shares the last direction plus two
transversal dimensions more, which will be named 78. These  two new
directions correspond to the two remaining compact dimensions of K3
that not belong to the vanishing 2-cycle.

 Note, that the D3-brane is effectively a D1-brane with an infinte mass, given that the two remaining directions,
parallel to the NS5-brane, are infinite. So, in the limit where the
two NS5 coincide, the D3-brane must be interpreted as an infinite
D1-brane pulling at both NS5-branes. Thus, the D3-brane is seen as a
point from the NS5-brane,  in a way that the nontrivial magnetic
configuration it carries is point-like in the (5+1)-dimensional
brane world. The fact that the effective D1-brane has infinite mass
is the reason why the Yang Monopole in our brane picture has
infinite energy, in agreement with the well-known gauge computation.
The use of the other transverse D1-brane, which sizes goes to zero,
is not other but to enhance the gauge symmetry.
\section{$SO(4)$-monopoles}\label{sec:so4}
Another D-brane construction can be achieved if we consider a $SO(4)$ gauge group
instead of $SU(2)$. This object will be called {\it extended}-Yang monopole~\cite{DL}.
The procedure is similar to the one shown in \ref{sec:SCOTYM}. Before entering into
details let us first notice that the existence of 2 charges in the usual Yang
construction is not obvious from the brane picture. D4-branes and D4-antibranes can each
wrap in two different ways a 2-cycle. The number of charges would be in principle 4,
each one labeling a possible realization. This is not the case as we are going to see.
Let us call
 \begin{equation*}
 \lambda_i:S^2\longrightarrow S^4 \qquad i=1,2
 \end{equation*}
  the two homotopically inequivalent maps of (\ref{eq:homo}), corresponding to
   the two ways  a  D$4$-brane wraps a 2-sphere, and $\tilde{\lambda}_i$ the
    homologous maps for a  D$4$-antibrane. The point is that there exists a
     homotopic deformation which makes $\lambda_1=\tilde{\lambda}_2$ and
      $\lambda_2=\tilde{\lambda}_1$, leaving us with just 2 (plus the trivial)
       homotopically inequivalent maps and, consequently, 2 charges. Roughly
        speaking, this means that the``one-way'' wrap of the D4-brane is actually
         identified with the ``other-way'' wrap of the D$4$-antibrane.

\begin{figure}
\begin{center}
 \includegraphics[width=5in]{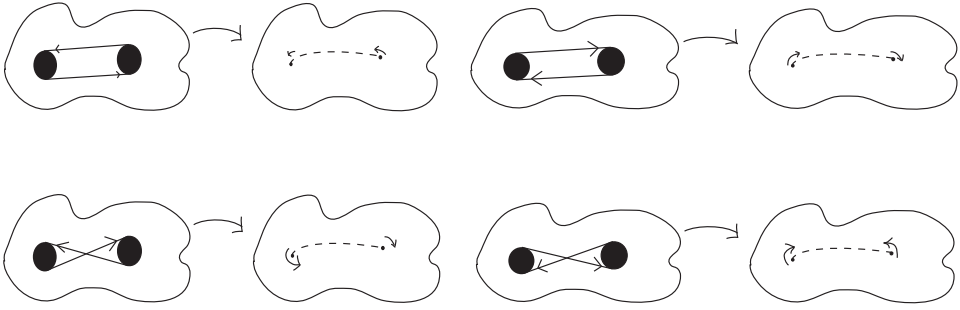}
\end{center}
\caption[$SO(4)$-monopole configurations]{{\small The four ways a
D4-brane can wrap two distinguishable vanishing 2-cycles of the  K3
surfcae. The configurations are homotopically different and so the
number of charges is four. Notice that at the singular limit (right)
the 2-dimensional part of the D4-brane connecting the two shrinking
cycles vanishes into a string and just the two wrapped cycles
remain.   }} \label{somonopoles}
\end{figure}

 Things are
  different if we consider an extended-Yang
  monopole. As shown in~\cite{DL}, imposing spherical
  symmetry to a $SO(4)$ bundle over $R^5-{0}$
  leaves four homotopically different possibilities.
  So the number of charges is 4
  in this case. This is intimately related to the fact that the
   algebra of $SO(4)$ is isomorphic to the Cartesian product of two  copies
    of $\mathfrak{su(2)}$, as can be visualized in
    its Dynkin diagram.  In the brane picture, the
    isomorphism of the algebras together with the
    geometric engineering mechanism we have
    used along subsection (\ref{sec:SCOTYM}), suggests
    that the construction of the $SO(4)$-monopole
    involves two vanishing 2-cycles on K3.
Recall that in the ALE space, each vanishing cycle is an $A_1$
singularity where the D2-brane is wrapped. Now, two D2-branes wrap a
shrinking cycle each, and geometrically engineer a
$\mathfrak{su(2)}$ factor. The singularities are well
separated\footnote{Strictly speaking, they would actually be at an
infinite distance in the ALE space.} and disconnected. They are
distinguishable. Now, as in the Yang case, we wrap them with a
D4-brane (see figure), which also wraps the rest\footnote{Think, for
example, that K3 occupies dimensions 6789. The 2-cycles may be both
placed at 67 and the D4-brane wraps it all.} of the K3 surface. As
before, the``one-way'' wrap of the D4-brane is identified with the
``other-way'' wrap of the D$4$-antibrane, so only branes are
considered. Now, the fact that $A_1$ singularities are
distinguishable rises the number of possible inequivalent
configurations to 4. The homotopy group which labels the
homotopically
        inequivalent maps is now
\begin{equation*}
\pi_4(A_1\times A_1)=\pi_4(A_1)\times\pi_4(A_1)=\mathbb{Z}_2\times\mathbb{Z}_2,
\end{equation*}
corresponding to the 4 charges for the extended-Yang monopole, in
agreement with~\cite{DL}.

\section{The multi-charge disease and the enhan\c con cure}\label{sec:MCDEC}
 As mentioned  in the introduction, there is an apparent
inconsistency in the D2-D4 system  on K3 model we have just considered and
which we claim as the Yang monopole. It precisely concerns
 its charges. There is a simple mechanism which would allow in
 principle to add an infinite
  charge to the monopole without breaking the  spherical symmetry.
  It would contradict the two
   charge nature of the Yang monopole and, consequently, the model would be incomplete.
   This is not the case though. We devote this  section to this issue.
\subsection{The problem}
The D4-D2 brane model proposed in~\cite{BDS2} seemed to fit the Yang
monopole gauge solution~\cite{Y} nicely: the dimension was correct;
the $SU(2)$ gauge group was geometrically engineered by the
D2-brane; the magnetic charge, carried by the D4-brane, had just two
possible configurations in accordance with the the two charges the
Yang solution permits; spherical symmetry was manifest by
construction; the model happened to be dual to another~\cite{BGT}
who claimed to have given a realization of the Yang monopole in
M-theory... However, as compared to the well-studied gauge
solutions, there was a crack in our brane model concerning the
multiple charge configurations and the (apparently) unavoidable
spherical symmetry the model showed. It has been objected that the
model was not by itself capable to
 explain the two-charge quality of the Yang monopole as opposed to the Z-tower of
 charges that characterizes Dirac monopoles\footnote{ We thank P. K. Townsend
 for this remark.}. The objection takes into account that there is a simple
 mechanism which would allow in principle to add an infinite
  charge to the monopole without breaking spherical symmetry. It soon urged
  for a deeper study since it would contradict the two
   charge nature of the Yang monopole and, consequently, the model proposed
   in ~\cite{BDS2} would be incomplete.

Let us go a little deeper in the problem and  explain how the model
cures itself. Consider a D4-brane wrapped on the K3 surface  and
located at $r=0$ in a polar coordination
 of the six dimensional spacetime. Being BPS states there
 is no reason against the addition of a second, a third or subsequent D4-brane superpositions.
 It results in a pile of an arbitrary number of D4-branes wrapped on the  K3 surface  at the same
  location, say $r=0$. Each brane accounts for a unit of magnetic charge, so the total
   charge is basically the number of D4-branes which is, in principle, arbitrary.
   Besides, the configuration manifestly preserves spherical symmetry from the six
    dimensional spacetime point of view since all the branes are located at the same point.
     In essence, this simple mechanism allows us to construct a monopole with arbitrary
     charge from our model which, for this reason, could no longer be claimed as the Yang
     monopole. Is there any way out?

Before answering the question, let us analyze the argument a little
deeper. In the original
 paper\cite{Y}, Yang did not merely construct the field strengths for the
 monopoles but also, and by means of the construction itself, proved that
 (up to isomorphism) there are only two nontrivial static spherically
 symmetric bundles of $SU(2)$ over $R^{5,1}-\{0\}$. They are respectively
 labeled by $\{+1,-1\}$ charges\footnote{A systematic analysis of the possible
 static spherically symmetric $SO(2n)$-bundles over $R^{2n+1,1}-\{0\}$,
 and consequently, of the charges that label them, is carried out in\cite{DL}.}.
 Now, it seems inappropriate to claim the model as the Yang monopole when its
 internal logic (the brane superposition) allows to add arbitrary charge.
 The field configurations they correspond to are either not $SU(2)$-bundles
 or they are simply not spherically symmetric. However, both properties have
 been explicitly and intentionally imposed in the construction of our
 brane model. {\it That} is the problem.

The enhan\c con mechanism (or more precisely, the fuzzy geometry it
involves for multicharge configurations), by means of which  the
second and subsequent branes feel a repulsive interaction with
respect to the first one and
 cannot reach the origin but smear onto the enhan\c con locus\footnote{The enhan\c con
 is, by definition, the locus of points where
 a probe brane gets tensionless in certain brane configurations as it tries to reach
 the origin.}at radius $r_e$, comes as a surprising solution for our problem. It is also a non
trivial prove of consistency of the brane model we propose for the
Yang monopole, a pure stringy mechanism that comes for the brane
picture to agree with a purely gauge theory requirement. The enhan\c
con mechanism was originally found in D$p$/D$(p+4)$-branes wrapped
in regular K3 surface~\cite{JPP}, bu it also works for fractional
branes~\cite{FLM}.

\subsection{The enhan\c con}
 In the picture shown in section \ref{sec:TSCYM} of the D2-D4 system  on K3 model we did not make
any assumption about the energy of the brane states. It was not
necessary since our discussion was purely
 geometrical and did not need the description of an explicit six
 dimensional effective
  field theory. In the following discussion however it becomes essential
  for the branes to be BPS states. \par
 When originally proposed\cite{JPP},
 the authors
  realized that their new mechanism (the enhan\c con) could resolve a naked timelike
  singularity
  produced by a D$p$-brane wrapped in a 2-cycle of the K3 surface
  which was being called {\it
   repulson}\footnote{In the geometry of a repulson, a massive particle
   would naively feel a repulsive gravitational force by a potential
   which becomes infinite as the particle approaches a point at finite
   distance from the physical location of the branes.}. In order to
   investigate it, they used the 10-dimensional supergravity of the
   system  D2-D6 on K3 model although they showed that the same
   conclusions hold for any  D$p$-D$(p+4)$ on K3 model\footnote{D$p$-branes
   are in the solution for consistency since even if one does not put
   them by hand they virtually appear.}. The geometric locus of the enhan\c con
is independent of the model. Let us see how the enhan\c con comes up in our model, for
the case $p=0$.

We will use the D0/D4-branes on the  K3 surface, that is, without
including D2-branes, so the gauge theory is abelian in the low
energy approximation that follows. D2-branes are not relevant for
the phenomenon we are describing,  they do not ``see'' the enhan\c
con shell. This fact allows us to simplify the computation but still
trust the result as extended for the full model.

The simplest static supergravity solution
consistently truncated to its bosonic part can be written as:
\begin{eqnarray}\label{sugra}
ds^2&=-&Z_0^{-1/2}Z_4^{-1/2}dt^2+ Z_0^{1/2}Z_4^{1/2}(dr^2+r^2d\Omega_4^2)+V^{1/2}Z_0^{1/2}Z_4^{-1/2}ds_{K3}^2,\nonumber \\
e^{2\Phi}&=&g_sZ_0^{3/2}Z_4^{-1/2}, \nonumber \\
C_{(1)}&=&(Z_0g_s)^{-1}dt, \nonumber \\
C_{(5)}&=&(Z_4g_s)^{-1}dt \wedge \epsilon_{K3}.
\end{eqnarray}
The line element corresponds to the string frame. $dS_{K3}^2$ is the
metric of the  K3 surface of unit volume, and $\epsilon_{K3}$ is its
corresponding volume form. Providing that the solution is
asymptotically flat, the harmonic functions are:
\begin{eqnarray}\label{harmonic}
Z_0&=&1- \frac{V_{*}}{V}f(r),\nonumber \\
Z_4&=&1+f(r),
\end{eqnarray}
where $V$ is the volume of K3 at $r=\infty$ and $V_{*}=(4\pi l_s)^4=\frac{\mu_0}{\mu_4}$. The
volume of K3 at arbitrary $r$ can be read
off from (\ref{sugra}):
\begin{equation}\label{V}
V(r)=V\frac{Z_0(r)}{Z_4(r)}.
\end{equation}
And the fuction $f(r)$ is, for a solution with $N$ branes:
\begin{equation}\label{f}
f(r)=\frac{1}{4}\frac{Ng_s l_s^3}{r^3}=\frac{V}{V_{*}}\frac{r_0^3}{r^3},
\end{equation}
where $r_0$ is the radius where a naked singularity (repulson) is
placed, as confirmed by inserting (\ref{f}) in (\ref{sugra}) and
computing the Kretschmann curvature scalar. As argued in~\cite{JPP},
this singularity is unphysical. This may be seen by probing the
geometry (\ref{sugra}) with other D4-brane. The action of the probe
may be written as
\begin{equation}\label{probe}
S_{\text{probe}}=\int{dt e^{-\Phi}\underbrace{(\mu_4 V(r)-\mu_0)}_{m(r)}\sqrt{-g}}+\mu_4\int{C_5}-\mu_0 \int{C_1},
\end{equation}
where the function $m(r)$ of the DBI term is the mass of the probe. The probe becomes massless
at $m(r)=0$, that is when $V(r)=V_{*}$.
We will define $r_e$ as the point where the K3 volume
becomes $V_{*}$, so $V(r_e)=V_{*}$.  Now, $r_e$ defines the enhan\c con locus. Its value
can be easily computed:
\begin{equation}\label{re}
f(r)=\frac{1}{2}(\frac{V}{V_{*}}-1)\longrightarrow r_e=\frac{2V}{V-V_{*}}|r_0|>|r_0|.
\end{equation}
The last inequality of (\ref{re}) shows that the enhan\c con radius is always bigger than $r_0$, the radius where the
naked singularity is placed, provided that $V>V_{*}$\footnote{The supergravity equations imply that $V(r)$ is an increasing function. So $V=V(r\to \infty)>V_{*}$ is always true.}.

  For smaller values of $r$ than the enhan\c con radius the tension of the probe
   becomes negative and the solution is considered unphysical
   from the supergravity perspective. The probe cannot approach
   the point $r=0$. Moreover, as ``seen'' by the probe,
   the $N$ D4-branes are not at $r=0$ but smeared  over
   the enhan\c con locus as well.



If we now try to start building the model, say, from empty space, the first
 D4-brane does not present any enhan\c con difficulties to reach $r=0$, so the
 1-(anti)brane solution is spherically symmetric. However, as we try to add a
  second and subsequent branes to the former, the enhan\c con mechanism
   prevents them to reach the first brane at the origin. Even if its not straightforward to
   see that this breaks spherical symmetry (the enhan\c con shell is, in principle, a sphere),
    it is not hard to believe that lack of point-like behavior will trigger multipole contributions.
    We will devote the next subsection to debate on this issue. For the time being let us accept it, and
     prelude that the model itself dynamically cures its apparent multi-charge contradiction.

There are two remarks that are worth pointed out. First, it should be noticed
 that the branes considered in seminal papers~\cite{JPP, JMPR, DPPR} were
 wrapping a nonvanishing 2-cycle of the K3 surface. Our model, by contrast, is built
  by the so-called fractional branes, that is, D$p$-branes which wrap vanishing
   2-cycles of the  K3 surface\footnote{D$p$-branes on  the  K3
   surface are equivalent to D($p-2$)-branes
   on $T^4/Z_2$, the orbifold limit of the K3 surface. So D4-branes becomes D2, and D2
    becomes D0 fractional branes.}.  This difference could raise some doubts
     on the above arguments. However, fractional brane solutions have also
      been proved to show enhan\c con behaviour\cite{FLM}. A second remark
       concerns about the gauge symmetry. As said before, although it is true
       that the explicit calculations were carried out only for an abelian gauge field,
       the extension for a $SU(2)$ Yang-Mills field (as geometrically engineered in
       our model) is not expected to show any obstruction for the enhan\c con given that
       the D2-brane is not affected by the enhan\c con.

\subsection{A way out of spherical symmetry}
As said some lines above is is not the enhan\c con mechanism but the fuzzy geometry it
involves what is behind the whole reasoning. Let us see that in detail.

In their first papers\cite{JPP,JMPR}, the authors considered the enhan\c con locus as
the spherical shell of radius $r_e$ where the stuck of $N$ branes become massless and
smeared out homogenously. So, even when the system itself prevents the branes from
piling at a six-dimensional point, say $r=0$, it seems that the solution is still
isotropic and then spherically symmetric. The original picture of the enhan\c con does
not actually break the spherical symmetry of the $N>1$-configuration of wrapped
D4-branes as claimed above.

It happens, however, that the homogenous distribution of $N$ branes over the spherical
enhan\c con shell is, as noticed in~\cite{AM, DJJ}, just a particular supergravity
solution. The enhan\c con, in general, has arbitrary shape. Think for example of an
oblate shape. One can then define a brane density on this enhan\c con surface whose
integral gives back the $N$ constituent branes. This density is not homogenous.
Moreover, as ruled by supergravity, for static solutions the brane density on the
enhan\c con behaves very much like an electrostatic distribution of charges on a
metal~\cite{AM}, that is, growing in regions where the curvature of the enhan\c con
surface is higher.

The spherical shape of the enhan\c con shell is then a particular case of a broad
family. The sphere was originally encountered not only for simplicity, but also because
the geometry was tested with a probe. Let us explain this point. The usual procedure for
the study of the enhan\c con geometry has been to take an initial set of $N$ gravitating
branes located at the origin. They are consequently a source of a spherically symmetric
geometry of the supergravity solution. Now, take the probe to test it. As the back
reaction of the probe is (by the definition of probe) not taken into account, there is
in principle no natural mechanism in the setup that might break spherical symmetry. So,
what was the point in considering different-from-spherical enhan\c con shells?

Consider the case of two monopoles. The first brane is placed at the origin so it is one
monopole. Now try to bring another {\it real} brane to wrap on it, another brane on the
same footing not a probe. The six dimensional supergravity ansatz for the metric of this
brane setup is clearly axially symmetric. So, what is the point in considering that the
branes finally melt out in a spherical enhan\c con shell?

The above argument suggests that there is in principle no especial
preference for spherical shells. Moreover, that a non spherical
geometry for the enhan\c con locus is naturally favoured. However,
although suggesting, the argument is inconclusive. Sphericalness may
seem capricious but still {\it possible} within the brane picture
whereas it is {\it forbidden} in a two or higher charge
configuration from the field theory point of view. A satisfactory
dynamical mechanism should rule out spherical enhan\c con shells.

Fortunately, there is an elegant way out the spherically symmetric
puzzle which involves phenomena beyond SUGRA. Recall that the
enhan\c con locus is the place where the probe brane (and also the
gravitating one as seen by the probe) becomes tensionless. At large
distance from $r=0$ the probe was pointlike in the six noncompact
dimensions. However as the distance to the center approaches $r_e$
the brane seems to ``emerge'' and smear on a four dimensional sphere
over the noncompact dimensions. Some lines above, it was explained
that the geometry of the enhan\c con need not to be spherical but a
large family of shapes are both available an consistent with the
enhan\c con mechanism~\cite{AM, DJJ}. This was a comforting way out
of the necessity of spherical symmetry but left us with an uneasy
family of arbitrary shapes which are all on equal footing with the
sphere. Why are all valid? Is there any mechanism that rules out any
of them? It seems a little naive though to think of such {\it sharp}
geometries at a region where the branes are blown in new
(noncompact) dimensions and spacetime seem not to behave ordinarily.
Indeed, it is believed that the correct description of the geometry
near the enhan\c con locus is a fuzzy sphere~\cite{M}. Fuzzy spheres
and more generally, non-commutative geometries break unavoidably
spherical symmetry. Moreover, Riemannian geometry and the concept of
manifold are no longer valid in this context. It is remarkable,
besides, that the fuzzy geometry appears for $N>1$ and so the charge-1
monopole does not get affected and recovers spherical symmetry as
expected. We will indicate within the next sections the relation
between the brane setup, the monopoles and the fuzzy geometry.
\section{ Connection with fuzzy spheres}

Fuzzy spheres are examples of noncommutative geometry~\cite{M}.
The main idea that underlies the construction of these spaces is
the one-to-one correspondence between the differential geometry
of manifolds and the commutative algebra of functions defined on
them. The coordinates are the generators of the algebra and the
vector fields are the derivations. A fuzzy sphere differs from
an ordinary sphere because the algebra of functions on it is not
commutative. Taking spherical harmonics as a basis, fuzzy spheres
are generated by harmonics whose spin $l$ is not greater than a
given $j$. The terms in the product of two spherical harmonics
that involves harmonics with exceeding $j$ are just omitted.
This truncation makes the algebra of functions noncommutative.
So fuzzy spheres are labeled by an integer number $j$. For $j=2$,
the algebra describes poorly a $S^2$-sphere, in fact, only the
north and the south pole are distinguishable. For $j=3$ one can
make out the equator as well, so the geometry gets less fuzzy.
The ordinary sphere is recovered in the limit $j\to \infty$.
In what follows we will turn $j$ into $N$, the number of branes,
because as we will see, the number of branes will actually rule
the degree of fuzziness. The radius of the fuzzy sphere is given by
\begin{equation}\label{eq:fuzzy}
R^2=\kappa (N^2-1),
\end{equation}
where $\kappa$ must be proportional to $\alpha'$ for dimensions.

In which sense are these fuzzy geometries related to the multiple
charge configurations? Unfortunately, we do not know the proper
mathematical framework to show how fuzzy geometry comes up in the
world volume of the NS5-branes in our model, as nicely shown
in~\cite{CMT} (by means of exploiting the DBI action) for non
abelian and non BPS setups where a pile of $N$ D1-branes stretch
between two D5-branes. We lack a low energy action for the
worldvolume of NS5-branes. However, it is commonly believed that the
enhan\c con mechanism comes along with fuzzy geometry, and in the
case of fuzzy spheres, (\ref{eq:fuzzy}) holds for the fuzzy radius,
where $N$ is the number of transversal branes. It is clear from
(\ref{eq:fuzzy}) that $N=1$ do not lead to fuzziness, but for any
$N>1$ it does. So multicharge solutions are fuzzy.

There is a remarkably nice explanation of how fuzzy geometry comes out from a similar setup~\cite{JPP} to our model.
It is worth reviewing it in this section.


 The model they used was built with D6-D2 brane system wrapped
on the K3 surface. In previous works~\cite{JMPR}, the authors already noticed that
such system would make a t' Hooft-Polyakov monopole when dimensionally reduced to four
dimensions. Specifically, they consider a D6-brane wrapping the  K3 surface and two more
compact flat dimensions, where the D2-brane lives. The enhan\c con mechanism produces an
enhancement of the gauge symmetry $U(1) \to SU(2)$ in the 4-dimensional region $r<r_e$,
which will be the core of the monopole. The configuration gets Higgsed as one moves far
from $r_e$ recovering the $U(1)$-magnetic charge.

In order to better visualize it, the brane setup that was actually
used~\cite{J,J2} to show how fuzzy spacetime geometry enters in the
enhan\c con picture was not the D6-D2 system  on K3 model already
mentioned but a dual one in Type IIB instead, which consists of a
pair of parallel NS5-branes with $N$ D3-branes stretched between
them. The separation of the NS5-branes is parameterized by
$\sigma\in[-1,+1]$ at which extremes the branes are initially
located. The presence of the D3-branes deforms the geometry of the
NS5-branes into a double trumpet shape~\cite{J}. The enhcan\c con
locus is precisely a transversal section of the bunch of D3-branes
at the point (say $\sigma=0$) where the two NS5-branes make contact.
As a consequence of the two brane connection, the gauge symmetry
gets enhanced to SU(2) at the enhan\c con locus. That was expected.
It is also expected that $N$ BPS monopoles enter this picture. They
are placed at the ends of the D3-branes as seen by the part of the
NS5-brane worldvolume transverse to the D3-brane. The positions of
those ends can be coordinated by $\Phi^i$, ($i=1,2,3$). This
``coordinates'' fulfill Nahm equations in the BPS case:
\begin{equation}\label{nahm}
\frac{d\Phi^i}{d\sigma}=\frac{1}{2}\epsilon_{ijk}[\Phi^j,\Phi^k].
\end{equation}
The appropriate solutions (BPS monopoles) are those for which the
$\Phi^i(\sigma)$ have a single pole at the end of the interval
$\sigma\in [-1,+1]$ and the residues $\Sigma^i$ form the $N\times N$
irreducible representation of $SU(2)$: $[\Sigma^i,\Sigma^j]=2i
\epsilon_{ijk}\Sigma^k$. The general solution of this kind takes one
$SU(2)$ representation and twists it to another as it crosses the
interval. The enhan\c con (recall, a section on the stuck of
D3-branes) is a fuzzy sphere for finite $N\neq 1$~\cite{M}. For
example, for a two D3-brane setup, a section will capture only two
clear points of $S^2$, the north and south pole. For a single
D3-brane, however, the radius of the fuzzy sphere vanishes
(\ref{eq:fuzzy}). For large $N$ the geometry becomes less fuzzy,
recovering the usual $S^2$ as $N\to \infty$.

 Their construction is different from the D4-D2 system  on the K3 surface we propose for
the six dimensional Yang monopole. Differences include, for
instance, the origin of the enhanced $SU(2)$ symmetry. We did not
make use of the enhan\c con mechanism but we geometrically
engineered the gauge group instead, as explained in section
\ref{sec:TSCYM}. We did so in order to account for a $SU(2)$ gauge
group all over the six dimensional space (and not only within the
enhan\c con region), as needed for the Yang monopole configuration.
Unfortunately, for the case D1-branes stretched between a pair
NS5-brane (our T-dual model), we lack the Nahm equations and an
analogous procedure as just shown for the emergence of fuzzy spheres
cannot be set. It would be very interesting to study the analogous
algebra in our model for the emergence of the fuzzy 4-sphere at the
enhan\c con locus. We leave it for a future work.

Despite this and other obvious differences the essence of the
problem concerning spherical symmetry remains. It is because the
enhan\c con shell they considered was spherical and supported
homogenously a melting $N$ D6-branes on it. An inconsistency with
field theory is again encountered (and cured by the fuzzy geometry)
in their case. That is because spherically symmetric t'
Hooft-Polyakov monopoles with multiple magnetic charge do not
exist~\cite{WG}.

Needles to say that the same enhan\c con-fuzzy mechanism applies for the $SO(4)$
monopole and cures it from the multicharge disease which could undergo in the
brane picture had the geometry not become fuzzy.

\section{Discussion and open questions}
In this  paper, a  Type IIA geometric  realization of the  Yang
monopole in six dimensions  given in~\cite{BDS2} is revisited and its
apparent contradictions are
 clarified. In the construction of the magnetic object, it has been   used  the
 result of the duality  between
 Type IIA superstring    compactified on    the   K3 surface  and    heterotic superstring
   on  $T^4$.   The $SU(2)$ gauge symmetry of the Yang monopole
 is considered as the  enhanced gauge symmetry  corresponding
 to  shrinking 2-cycles inside
   the  K3 surface, and  the  Yang monopole  comes up by
   wrapping D-branes on the K3 non-trivial cycles.
    In this way, the properties of the Yang monopole are  encoded in the K3 surface data.

With respect to the charges of the configuration, suggestions and objections that came up
 during the presentation of~\cite{BDS2} have been taking in full consideration. In our
  opinion, the present work brings light to the main objections strengthening and completing
   the brane picture of the Yang monopole. Firstly, it was claimed that the number of charges
   of one D4-brane setup should be four, as accounted
     for the two ways the brane and the antibrane can wrap a 2-cycle. The answer is given at
      the end of section \ref{sec:TSCYM}. There it is explained that the four configurations
       are actually identified in pairs, so it results in just two homotopically different
        configurations. Indeed, as explain in the same section, the $SO(4)$ extended-Yang
         monopole is the one who carries four charges and a brane picture for it is proposed.

The second objection have been taken into analysis in section
 \ref{sec:MCDEC}. When more than one D4-brane are added to the model, its interpretation
 as a Yang monopole gets into trouble since and infinite tower of charges seem to appear.
 This is what we have called multicharge disease. The multi-charge problem of this
  construction gets satisfactory solved by the dynamics of the enhan\c con mechanism
   which, as explained in section \ref{sec:MCDEC}, ruins spherical symmetry in the
   multi-brane setup and then saves the model from contradiction.

The question of how such an involved concept as the enhan\c con locus and its correlated
fuzzy geometry comes into stage to make the brane configuration non spherically
symmetric as required, for different reasons, by the gauge field theory is a point that
in our opinion requires further analysis. More on this will be reported in future works.



\section*{Acknowledgements} \nonumber
We thank   M. Asorey, L. J. Boya, E. H. Saidi,  P.  K. Townsend, K.
Goldstein and C. Hoyos-Badajoz for  discussions and enlightening
comments. This work has been partially supported by CICYT (grant
FPA-2009-09638) and DGIID-DGA (grant 2007-E24/2). We thank also the
support by grant A9335/07 and A9335/10 (Fisica de alta energia:
Particulas, cuerdas y cosmologia).

\end{document}